\documentstyle[psfig,epsfig,aps,prd]{revtex}

\def\beq{\begin{equation}}
\def\eeq{\end{equation}}
\def\lsim{\raise0.3ex\hbox{$\;<$\kern-0.75em\raise-1.1ex\hbox{$\sim\;$}}}
\def\gsim{\raise0.3ex\hbox{$\;>$\kern-0.75em\raise-1.1ex\hbox{$\sim\;$}}}

\def\aj#1#2#3{ Astron. J. {\bf #1}, #3 (19#2)}
\def\apj#1#2#3{ Astrophys. J. {\bf #1}, #3 (19#2)}

\def\mn#1#2#3{ MNRAS {\bf #1}, #3 (19#2)}
 
\def\np#1#2#3{ Nucl. Phys. {\bf #1}, #3 (19#2)}
\def\nat#1#2#3{ Nature {\bf #1}, #3 (19#2)}

\def\pr#1#2#3{ Phys. Rev. {\bf #1}, #3 (19#2)}

\def\prl#1#2#3{ Phys. Rev. Lett. {\bf #1}, #3 (19#2)}

\begin{document}
\twocolumn[\hsize\textwidth\columnwidth\hsize\csname
@twocolumnfalse\endcsname
\title{Remarks on the Boomerang results, the baryon density and the
leptonic asymmetry}
\author{Julien Lesgourgues and Marco Peloso}
\address{SISSA--ISAS and INFN, Sezione di Trieste,
Via Beirut 2-4,I-34013 Trieste, ITALY \\
E-mail: lesgour@sissa.it, peloso@sissa.it}
\date{Preprint sent 28 April 2000; submitted 1st May 2000}
\maketitle

\begin{abstract}
The very recent Boomerang results give an estimate of unprecedented
precision of the Cosmic Microwave Background anisotropies on
sub--degree scales. A puzzling feature for theoretical cosmology is
the low amplitude of the second acoustic peak.  Through a qualitative
discussion, we argue that a scarcely considered category of flat
models, with a leptonic asymmetry, a high baryon density and a low
cosmological constant seems to be in very good agreement with the
data, while still being compatible with big bang nucleosynthesis and
some other observational constraints.  Although this is certainly not
the only way to explain the data, we believe that these models deserve
to be included in forthcoming likelihood analyses.
\end{abstract}


\vskip2pc]

\section{Introduction}

The measurement of Cosmic Microwave Background (CMB) anisotropies has
been driving the attention of cosmologists over the past decade.  Very
recently, De Bernardis et al. \cite{boomerang} published the first results of
the Boomerang balloon Antarctic flight. With these data, the
(recent) story of precision cosmology climbs a new step. For the first
time, the anisotropy power spectrum has been measured by a single
experiment in a wide range of angular scales, from multipoles
$l \sim 50$ up to $l \sim 600$, with many independent points, and
error bars of order $\pm$ 20 percent. The observation of a narrow peak,
centered around multipoles $l \simeq 200$, confirms the inflationary
picture of an approximately flat Universe with adiabatic fluctuations.

This beautiful result was already suggested by previous CMB
experiments (see \cite{ddbb} for a recent review). On the other hand,
the Boomerang anisotropy spectrum exhibits a puzzling behavior on
small scales (high multipoles): in the range in which a pronounced
secondary peak was expected, the data points are rather low, with an
almost flat shape. It seems that the cosmological model most favored
during the past year, which is a flat Cold Dark Matter (CDM) model with a
large cosmological constant and ``standard'' parameter values (see
below), can hardly account for this feature, unless some new ingredient is
added.

After this paper was submitted, a detailed analysis of the data by the
Boomerang team was released in ref.\cite{Lange}, followed by another
work \cite{Balbi} including also the new MAXIMA data \cite{Hanany}. We
refer the reader to these works for a more exhaustive
interpretation of this puzzling small--scale behavior. In the
following discussion, we just intend to point out that a particular
category of models, which are scarcely taken into account, seem to be
in remarkable agreement with the new published data (as can be seen
from Fig.1). We therefore believe that they deserve some attention,
and should be included in future data analyses.

While this discussion was being completed, a nice paper by White et
al. \cite{Elena} was put on the preprint database, suggesting many
possible explanations of the data, including a large baryon density
$\Omega_b$.\footnote{In addition, the possibility of explaining the
data with a high $\Omega_b$ was confirmed contemporaneously
\cite{Tegmark} and soon after \cite{Lange,Balbi} the release of this work.} 
The key point of this rapid communication is to
recall that a large $\Omega_b$ is still in agreement with the observed
light element abundances, provided that a large neutrino asymmetry is also
present \cite{Kang}. With these two ingredients (large $\Omega_b$ and
neutrino asymmetry) the Boomerang data can be nicely fitted in a flat
Universe context, especially with a low value of the cosmological
constant. Interestingly, this class of
models can satisfy some other cosmological requirements, such as the
ones coming from the matter power spectrum and from the baryon
fraction in clusters.

\section{Flat $\Lambda$CDM models and Boomerang data}

As a starting point, we plot on Fig.1 (solid line) a $\Lambda$CDM
model with the parameters which were recently the most favored:
$(\Omega_{tot}, \Omega_{\Lambda}, h^2 \Omega_b, h, n) = (1,0.70,
0.019, 0.68, 1)$, with no tensors and reionization neglected ($h$ is
the Hubble constant in units of $100 \,{\mathrm km}\, {\mathrm
s}^{-\,1} \, {\mathrm Mpc}^{-\,1}$, $n$ stands for the scalar
primordial spectrum tilt).  This set is in agreement with standard BBN
\cite{Tytler,Olive}, supernovae \cite{Perlmutter},
clusters\cite{Bahcall}, direct measurements of $h$ \cite{younger},
constraints on the matter power spectrum \cite{kofman}, and other
observations. However, as we can see from Fig.1, once the
Boomerang data are taken into account, the first and second peaks do
not have the right shape and location. The problem is not so much with
the overall amplitude of the peaks, which can be affected by calibration 
uncertainties, and which
can be adjusted by changing
the scalar tilt, the reionization optical depth $\tau$ and the tensor
amount. Most difficult is to accomplish the location, relative
amplitude and shape of the peaks, which depend only slightly on the
previous parameters. Rather, we
must tune other parameters in order to shift the peaks to the left
(smaller multipoles), and flatten the second peak.  So, if we remain
in the framework of a flat $\Lambda$CDM model with power--law
primordial spectrum, the only way to get closer to the experimental
data is by changing $h^2 \Omega_b$, $\Omega_{\Lambda}$ or $h$.

For instance, by lowering $h^2 \Omega_b$, we could shift the peaks to
the left. But a low baryon density would enhance even peaks with
respect to odd ones. Clearly, this is not favored by the data: the
large ratio between the first and second peak amplitude rather
suggests a large baryon fraction. 
It seems that there are not many solution to this interesting and new
situation (at least if we avoid introducing a positive curvature or
some ``exotic'' cosmological parameters): it is necessary, first, to
decrease $\Omega_{\Lambda}$ and/or increase $h$ in order to have the
first peak on the right scale, and, second, to take a high baryon
density $h^2 \Omega_b$, in order to suppress the second peak amplitude
with respect to the first one. Then, the remaining parameters (scalar
tilt, overall normalization, tensor amount, reionization) can enter
into the game in order to adjust the overall peak amplitude with respect to the
Sachs--Wolfe plateau (only in this last stage the uncertainty on Boomerang
calibration and on COBE normalization play a role).

So, the Boomerang result is so characteristic (with its low and flat
second peak) that even without a precise likelihood analysis, a quick
glance brings evidence for a large baryon density $h^2 \Omega_b>0.02$,
together with $h>0.7$ and/or $\Omega_{\Lambda} < 0.70$ (again, in the
framework of flat $\Lambda$CDM models). It is intriguing that previous
precise analyses of CMB data, which did not include the new Boomerang
results and the related information concerning the shape of the
secondary peak, also favored a high baryon density and a low
cosmological constant \cite{ddbb}. After this communication was put on
the database, the quantitative analyses based on Boomerang
\cite{Lange} and Boomerang + Maxima \cite{Balbi} results also pointed
towards a high baryon density.

\section{Including BBN constraints}

The large baryon density suggested by the previous analysis conflicts
current estimates from standard BBN, which indicate $h^2 \Omega_b
\simeq 0.019$, with an error bar varying between $\pm 0.001$ and $\pm
0.004$ in the recent literature \cite{Tytler,Olive}.

As pointed out for example by Kang and Steigman in 92 \cite{Kang}, a
high value for $h^2 \Omega_b$ can still lead to the observed light
element abundances, provided there is a large asymmetry between
neutrinos and antineutrinos in the primordial plasma (degenerate
neutrinos).\footnote{BBN bounds the number of baryon minus antibaryons
to be very small with respect to the number of photons. Since the
Universe appears to be electrically neutral, an asymmetry on the
charged leptons is equally bounded to be vanishingly small. However,
the possibility of a large asymmetry in the neutrino sector cannot be
excluded. From a particle physics point of view, a large lepton
asymmetry can be generated by an Affleck--Dine mechanism \cite{AF}
without producing a large baryon asymmetry (see
refs. \cite{Casas}), or even by active--sterile neutrino
oscillations \cite{Foot}. In general, the asymmetry is
different for each neutrino family.}  One of the main effects of this
asymmetry is to enhance the relativistic energy density, usually
parametrized by the number $N_{{\mathrm eff}}$ of effective
relativistic neutrino families, with a consequent increase of the
expansion rate of the Universe. If $\mu_i$ and $T_i$ denote,
respectively, the chemical potential and the temperature of the $i-$th
neutrino family ($ i = e,\mu, \tau$), this effective number is linked
to the degeneracy parameters $\xi_i \equiv \mu_i / T_i $ by:
\begin{equation}
N_{{\mathrm eff}} = 3 + \Sigma_i \left[ \frac{15}{7} \left(
\frac{\xi_i}{\pi} \right)^4 + \frac{30}{7} \left( \frac{\xi_i}{\pi}
\right)^2 \right] \;\;.
\end{equation}
For what concerns BBN, $N_{\mathrm eff}>3$ leads to a higher neutron
to proton ratio, since $n$, $p$ decouple earlier from the primordial
plasma. On the contrary, the presence of a positive asymmetry for the
electronic neutrinos ($\xi_e > 0$) tends to reduce this ratio, since
it ``shifts towards the proton direction'' the reaction $n \, \nu_e
\leftrightarrow p \, e^-$ (and the crossed ones). The two effects can
compensate each other in a wide region in the ($\xi_e$,
$\xi_{\mu,\tau}$) parameter space, and the observed abundances of
light elements can be achieved with a value of $h^2 \Omega_b$
significantly higher than the bound coming from standard BBN ($\xi_i =
0$).

For a quantitative analysis, one can apply the results of \cite{Kang},
which associate to any given value of the baryon density a region in
the ($\xi_e$, $N_{{\mathrm eff}}$) plane where primordial
nucleosynthesis is successful. In this work, the lower and upper
bounds on $N_{{\mathrm eff}}$ come, respectively, from the requirement
that $^7$Li is not too abundant ($^7{\mathrm Li}/{\mathrm H} \geq 3
\cdot 10^{-\,5}$) and that enough deuterium is produced (actually
$\left( {\mathrm D} + {}^3{\mathrm He} \right)/{\mathrm H} \leq
10^{-4}$). The most stringent limits on $\xi_e$ come instead from the
observed $^4$He abundance (in \cite{Kang} the helium$-4$ mass fraction
$Y$ is assumed to be in the conservative range $0.21-0.25\,$). While
more recent observations put more severe bounds on this last quantity,
the above estimates on lithium--7 and deuterium abundances are in good
agreement with the latest ones (see for example \cite{Tytler} for a
review). As a consequence, the allowed region in ($h^2 \Omega_b,
N_{\mathrm eff}$) given by \cite{Kang} remains in good agreement with
recent data (see also \cite{Napoli} for an updated analysis).

Cosmological implications of ``degenerate BBN'' have been the object
of several recent studies.  The importance of a large leptonic
asymmetry for the formation of large--scale structure (resp. CMB
anisotropies) was pointed out by \cite{Larsen} (resp. \cite{Sarkar}), but
the first comparison of $\Lambda$CDM models with both CMB and large--scale
structure data was performed in \cite{LP}. Another question was
addressed by Kinney and Riotto \cite{Kinney}, who calculated the
sensitivity of forthcoming CMB satellites to the neutrino degeneracy
parameter. The analysis was extended to the degenerate neutrino mass
in \cite{LP2}. We should also stress that ref. \cite{Gelmini} proposed
a lepton asymmetry for generating ultra--high energy cosmic rays beyond
the Greisen--Zatsepin--Kuzmin cut--off.

In ref. \cite{LP}, a $\Lambda$CDM model with large leptonic asymmetry
was compared with the CMB data available at that time, plus a few
constraints from large--scale structure, the most restrictive being
the matter power spectrum normalization to $\sigma_8$ (the variance of
mass fluctuations in a sphere of radius $R=8 h^{-1}$Mpc). The effect
of massless degenerate neutrinos is mainly to postpone the time of
equality, therefore boosting the first acoustic peak, shifting the
peaks to higher multipoles, and suppressing small--scale matter
fluctuations. It can be completely modeled with the effective neutrino
number $N_{\mathrm eff}$ introduced above, in contrast with the case
of massive degenerate neutrinos (for which simple modifications to
{\sc cmbfast} \cite{SelZal} must be performed). It was shown in
\cite{LP} that high values of the cosmological constant (such as
$\Omega_{\Lambda}\geq 0.80$) are hardly compatible with $N_{\mathrm
eff} > 3$, while for $\Omega_{\Lambda} \leq 0.70$ there are some
allowed windows in the space of cosmological parameters, ranging up to
very large effective neutrino numbers. An interesting point is that in
refs.\cite{Sarkar,LP}, even with $\Omega_{\Lambda}=0$, an agreement
was found with both CMB and large--scale structure constraints.

It is amazing that Boomerang data seems precisely to favor, as one of
the simplest possibilities, a high baryon density combined with a low
cosmological constant. The high baryon density requires a leptonic
asymmetry in order to be compatible with the observed light elements
abundances (unless systematic errors have been underestimated in all
recent analyses), and following ref. \cite{LP}, this large asymmetry
is allowed precisely in presence of a low cosmological constant. In
Fig.1 we provide some examples of such models (dashed and dotted lines).
\begin{figure} [htb]
\centerline{\psfig{file=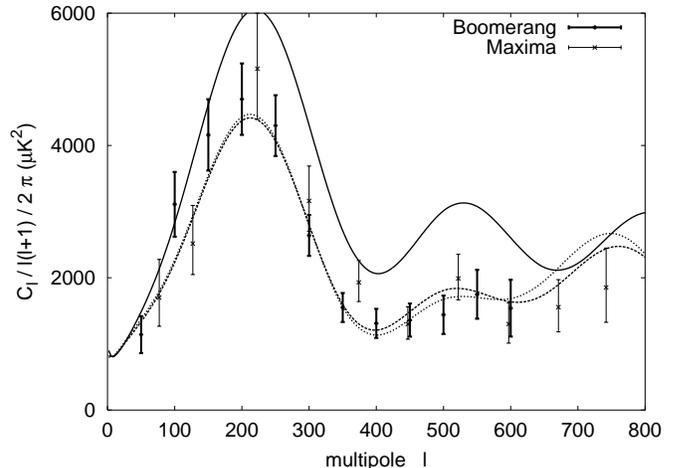,width=0.5\textwidth}}
\caption{ In solid line, a $\Lambda$CDM model (favored by many
pre--Boomerang experiments) with $(\Omega_{tot}, \Omega_{\Lambda}, h^2
\Omega_b, h, n)$ = $(1,0.70, 0.019, 0.68, 1)$, COBE normalization, no
tensors and reionization neglected. In dashed and dotted lines, two
models with a high baryon density compensated in the spirit of
``degenerate BBN'' by a large effective neutrino number. Parameters
are $(N_{\mathrm eff}, \Omega_{\Lambda}, h^2 \Omega_b, h, n, \tau)$ =
(6, 0.20, 0.025, 0.7, 0.98, 0.05) (dashed curve) and (6, 0, 0.028,
0.7, 1, 0.05) (dotted curve), with no tensors.  The Boomerang points
are taken from De Bernardis et al. [1], and we show only the error
bars accounting for noise and cosmic variance. We also add the MAXIMA
points from Hanany et al. [5] which were released after submission of
this paper, without altering our main conclusions (the calibration of
MAXIMA is ajusted relatively to the one of Boomerang as in ref. [5]).}
\end{figure}

\section{Compatibility with large--scale structure and other constraints}

There are several independent observations which suggest the presence
of a non-vanishing cosmological constant. Very recently, supernovae
data \cite{Perlmutter} motivated several works in this direction.
Also, indications for $\Omega_m < 1$ (and thus $\Omega_\Lambda > 0$ if
the Universe is assumed to be flat) are provided, for example, by the
study of matter abundance (baryons $+$ cold dark matter) in clusters,
by the limits on the age of the Universe, or by constraints on the
matter power spectrum (see \cite{triangle} for a recent review).  It
seems to be particularly convincing that most of these observations
favor a common result $\Omega_m \sim 0.2 - 0.3\,$. However, at
present, none of them can be said to be conclusive if considered
separately from the others. Before firmly relying on $\Omega_\Lambda
\geq 0.7$, it is thus legitimate to investigate if some of the
arguments listed above can be evaded.  For example, the model
considered in the previous section, with $\Omega_m > 0.3$, large $h^2
\Omega_b$ and neutrino degeneracy, is not excluded by constraints on
the observed matter power spectrum and on the fraction of baryonic
mass in clusters, which is one of the most robust arguments for a low
$\Omega_m$.

As far as the latter is concerned, the cosmological baryon density can
be constrained by the ratio of the baryonic mass to the total
gravitational mass in clusters \cite{White}. Numerical simulations show
that this ratio should be nearly equal (actually slightly lower) to
the cosmological average. Thus one can evaluate the ratio in clusters
(the baryon fraction in clusters can be deduced from X--ray emission,
while the total matter can be extracted from the velocity dispersion
curves) and gain a relation between the baryon and matter densities in
the Universe. Tytler el al. \cite{Tytler} report the following bound:
\begin{equation}
h^2 \Omega_b \geq \left( 0.025 - 0.060 \right) h_{70}^{1/2} 
\Omega_m \;\;,
\end{equation}
where $h_{70}$ is the Hubble constant in units of $70 \,{\mathrm km}\,
{\mathrm s}^{-\,1} \, {\mathrm Mpc}^{-\,1}$.  One usually assumes $h^2
\Omega_b$ to be bounded by the values allowed by standard BBN. As a
consequence, $\Omega_m$ has to be smaller than one. However, the
values for $h^2 \Omega_b$ considered in the previous sections are
compatible even with a Universe closed by matter alone.

Finally, $\Lambda$CDM models with a low (or vanishing) cosmological
constant are known to be hardly compatible with large--scale structure
data (what is also known as the ``shape parameter'' problem).  Indeed,
once the primordial spectrum has been normalized to COBE, there is an
excess of power in small--scale matter fluctuation with respect to the
bounds on $\sigma_8$. This issue cannot be solved by introducing a
large red tilt (then, the shape of the power spectrum contradicts
redshift surveys), neither with a large tensor amount (that would
suppress the CMB acoustic peaks). On the other hand, it was shown in
\cite{LP} that the neutrino degeneracy, by postponing matter
domination, and suppressing small--scale matter fluctuations, can
reconcile CMB data with large--scale structure constraints, even for a
zero cosmological constant (in other words, the ``shape parameter'' is
consistent with current estimates). The proof was made for particular
values of the cosmological parameters ($h$, the tensor amount and the
optical depth to reionization were not allowed to vary), but it
indicates clearly that in a systematic analysis, agreement with
COBE, Boomerang and other CMB data (including calibration
uncertainties) is not exclusive from a correct shape and amplitude for
the matter power spectrum.

\section{Conclusions}

In conclusion, we believe that models with a large neutrino asymmetry
deserve to be included in forthcoming precise comparisons with
experimental data. In practice, this amounts in including
simultaneously higher values of $N_{\mathrm eff}$ and $h^2 \Omega_b$
than the ones usually considered.

Since a high baryon density enhances odd peaks with respect to even
ones, a natural outcome of these models is a large amplitude for the
third peak. This will be probably the best way to test this scheme in
a near future. If the third peak turns out to be also very low, the
situation will be even more puzzling, and more complicated models (for
instance, with a Broken--Scale--Invariant primordial spectrum
\cite{LPS} or with topological defects \cite{Patrick}) may have to
enter into the game.
 
\section*{Acknowledgements}

We are grateful to Sergio Pastor for several useful comments.
The CMB power spectra are computed with {\sc cmbfast} by Seljak and
Zaldarriaga \cite{SelZal}. J.~L. is supported by INFN and by the TMR
network grant ERBFMRXCT960090.

\end{document}